\documentclass[aps,prr,twocolumn,groupedaddress,superscriptaddress]{revtex4-2}
\usepackage{graphicx}
\usepackage{amssymb,amsmath}
\usepackage{epstopdf}
\usepackage{float}
\usepackage{xcolor}
\usepackage{pifont}
\usepackage{bm}
\usepackage{mathrsfs}
\usepackage{bbold}
\usepackage{upgreek}
\usepackage{braket}

\usepackage{times}
\usepackage{url}
\usepackage{xcolor}
\usepackage{gensymb}

\begin{document}

\title{Observation of quantum-classical transition behavior of LGI in a dissipative quantum gas}


\author{Qinxuan Peng}
\thanks{These authors contributed equally to this work.}
\affiliation{School of Physics and Astronomy, Sun Yat-sen University, Zhuhai 519082, China}
\author{Bolong Jiao}
\thanks{These authors contributed equally to this work.}
\affiliation{School of Physics and Astronomy, Sun Yat-sen University, Zhuhai 519082, China}
\author{Hang Yu}
\affiliation{School of Physics and Astronomy, Sun Yat-sen University, Zhuhai 519082, China}
\author{Liao Sun}
\affiliation{School of Physics and Astronomy, Sun Yat-sen University, Zhuhai 519082, China}
\author{Haoyi Zhang}
\affiliation{School of Physics and Astronomy, Sun Yat-sen University, Zhuhai 519082, China}
\author{Jiaming Li}
\email[]{lijiam29@mail.sysu.edu.cn}
\affiliation{School of Physics and Astronomy, Sun Yat-sen University, Zhuhai 519082, China}
\affiliation{Shenzhen Research Institute of Sun Yat-Sen University, Shenzhen 518057, China}
\affiliation{State Key Laboratory of Optoelectronic Materials and Technologies, Sun Yat-Sen University, Guangzhou 510275, China}
\affiliation{Guangdong Provincial Key Laboratory of Quantum Metrology and Sensing, Sun Yat-Sen University, Zhuhai 519082, China}

\author{Le Luo}
\email[]{luole5@mail.sysu.edu.cn}
\affiliation{School of Physics and Astronomy, Sun Yat-sen University, Zhuhai 519082, China}
\affiliation{Shenzhen Research Institute of Sun Yat-Sen University, Shenzhen 518057, China}
\affiliation{State Key Laboratory of Optoelectronic Materials and Technologies, Sun Yat-Sen University, Guangzhou 510275, China}
\affiliation{Guangdong Provincial Key Laboratory of Quantum Metrology and Sensing, Sun Yat-Sen University, Zhuhai 519082, China}
\affiliation{Quantum Science Center of Guangdong-HongKong-Macao Greater Bay, Shenzhen 518048, China}

\date{\today}

\begin{abstract}

The Leggett-Garg inequality (LGI) is a powerful tool for distinguishing between quantum and classical properties in studies of macroscopic systems. Applying the LGI to non-Hermitian systems with dissipation presents a fascinating opportunity, as competing mechanisms can either strengthen or weaken LGI violations. On one hand, dissipation-induced nonlinear interactions amplify LGI violations compared to Hermitian systems; on the other hand, dissipation leads to decoherence, which could weaken the LGI violation. In this paper, we investigate a non-Hermitian system of ultracold Fermi gas with dissipation. Our experiments reveal that as dissipation increases, the upper bound of the third-order LGI parameter $K_3$
initially rises, reaching its maximum at the exceptional point (EP), where $K_3 = C_{21} + C_{32} - C_{31}$, encompassing three two-time correlation functions. Beyond a certain dissipation threshold, the LGI violation weakens, approaching the classical limit, indicating a quantum-to-classical transition (QCT). Furthermore, we observe that the LGI violation decreases with increasing evolution time, reinforcing the QCT in the time domain. This study provides a crucial stepping stone for using the LGI to explore the QCT in many-body open quantum systems.


\end{abstract}

\maketitle

\section{Introduction}

	


Since its inception in the last century, quantum mechanics has achieved tremendous success, revealing non-classical phenomena such as superposition and entanglement. However, measures aiming to quantify various aspects of macroscopic quantumness remains an open question worthy of investigation~\cite{Frowis2018}. One particular interesting topic is QCT that is elucidated through the orthodox Copenhagen interpretation of wave function collapse resulting from measurements~\cite{stapp1972,bassi2013}, and the transition is further comprehended through the quantum decoherence theory~\cite{zurek1991,haroche1998,schlosshauer2005,Schlosshauer2007,schlosshauer2019,zurek2003decoherence}. Quantum decoherence is often closely associated with macroscopic quantum systems, particularly those that interact with the environment. These open systems provide an ideal platform for exploring the boundary between the quantum world and classical reality, where the transition from quantum to classical is not only a topic of theoretical interests but also a practical concern for the applications of quantum technologies~\cite{adler2009,schlosshauer2005,zurek2009,carlesso2022,jeong2014,camilleri2015niels,paz2002environment}.

Open quantum systems are defined as composed of localized, microscopic regions coupled to the external environment through appropriate interactions~\cite{Rotter2015,szankowski2023introduction}. In these systems, the evolution of quantum states is influenced not only by the system's internal Hamiltonian but also by interactions with the environment. Open quantum systems have been realized in many systems, such as cold atoms~\cite{li2019}, trapped ions~\cite{bian2023,lu2023}, superconducting circuits~\cite{chen2022decoherence}, electric circuit~\cite{wang2020observation}, microcavities~\cite{wang2021}, and nitrogen-vacancy centers~\cite{wu2019}. The quantitative description of the time-dependent evolution of such systems interacting with the environment is often achieved through the Lindblad master equation. When the back interaction from the environment to the system is ignored, equivalent to the absence of the quantum jump term.  the open quantum system can be viewed as a non-Hermitian system described by Schrödinger equation with nonunitary time evolution operator.
 
 Non-Hermitian quantum systems are firstly realized in ultracold quantum gases through precisely controlled spin-dependent dissipation~\cite{li2019}. These systems have elucidated key mechanisms in complex dynamical processes, such as the impact of EP and parity-time ($\mathcal{PT}$) symmetry breaking~\cite{li2023unification}. Here we further suggest that experiment with a dissipative quantum gas could help to  understand the QCT behavior in a non-Hermitian many-body system, since dissipation is always accompanied by decoherence as a result of entanglement with the environment~\cite{schlosshauer2019quantum}. In these systems we could engineer  the loss of coherence many orders of magnitude faster than any relaxation processes induced by dissipation. This rapid decoherence allows to study the QCT in the non-Hermitian system much more efficiently compared to the Hermitian one. Additionally, non-Hermitian systems feature EP, where eigenvalues and eigenvectors coalesce. Approaching these points, the system could undergo a rapid dynamic transition, which may offer a unique setting to gain deep insights into the QCT.



The LGI serves as a fundamental framework for evaluating whether macroscopic systems have quantum properties~\cite{Xu2011,vitagliano2023leggett,ghosh2023leggett}. It relies on the assumptions of macrorealism (MR) and non-invasive measurement (NIM). Theoretically, LGI involves a series of measurements on a system at different times, functioning similarly to a temporal analog of the Bell inequalities. Violation of the LGI indicates that the system does not adhere to MR and NIM assumptions, thereby highlighting the presence of quantum effects. The LGI tests have been applied across various close quantum systems, including superconducting circuits~\cite{palacios2010}, single photon systems~\cite{zhou2015,dressel2011}, nuclear magnetic resonance systems~\cite{athalye2011,katiyar2013,katiyar2017experimental}, trapped ions~\cite{zhan2023}, and NV centers~\cite{waldherr2011,tusun2022}, but with little exploration on the role of the coupling between the system and environment. Recently, the breakthrough researches of testing the LGI in a non-Hermitian quantum system has been implemented with trapped ions~\cite{lu2023,wu2023maximizing,quinn2023observing}, where supercorrelation, the enhanced violation of LGI compared to the Hermitian quantum mechanics, have been observed in debt to interactions with the environment. But these researches are limited to a single qubit, where the decoherence effects of many-particle systems are absent. The physics of the QCT in a non-Hermitian system is yet to be explored, which requires a many-particle non-Hermitian ensembles.

In this study, we observe the QCT behavior using the LGI as a quantitative tool in an ultracold Fermi gases with spin dependent dissipation. Both dissipation and decoherence strength are controllable in this system. In the case of small decoherence strength, we observe an enhancement in the violation of the LGI that can be compared to the significant results discovered in Ref.~\cite{lu2023}. As the dissipation strength increases, approaching the EP, the LGI violation gradually weakens, indicating a QCT. When keeping the dissipation strength constant, an increase in the evolution time also leads to a gradual weakening of the LGI violation, reinforcing a QCT in the time domain. Thus, we suggest that the degree of LGI violation can be used as a quantitative tool to define the boundary between quantum and classical regimes in a non-Hermitian system.


\section{experiment method}

As shown in Fig.1(a) and (b), we prepare a non-interacting Fermi gas of $^{6}$Li at the two lowest $^{2}S_{1/2}$ hyperfine states ($\ket{F=1/2, m_F=1/2}$ and $\ket{F=1/2, m_F=-1/2}$), labeled as $\ket{0}$ and $\ket{1}$, respectively. These two-level states are coupled by a radio-frequency (RF) field with a coupling strength of $J$. A resonant optical beam is used to excite the atoms from $\ket{1}$ to the $^{2}P_{3/2}$ state denoted as $\ket{a}$ and generates atomic loss in $\ket{1}$ with the atom number of state $\ket{0}$ keeping constant~\cite{li2019}. The loss rate is defined as $\gamma$. Then the Hamiltonian of this dissipative two-level system is shown as:
\begin{equation}
	H=J \sigma_{\mathrm{x}}-i \gamma| 1 \rangle\langle 1 |=-i \mathbf{I} \gamma/ 2 +H_{\mathcal{PT}}
\end{equation}
where $\mathbf{I}$ is the unit matrix, and $H_{\mathcal{PT}}=J \sigma_x+i \sigma_z \gamma/2$ is a $\mathcal{PT}$-symmetric Hamiltonian. The eigenvalues of $H_{\mathcal{PT}}$ are $\pm \sqrt{J^{2}-\gamma^{2}/4}$. The EP is at $\gamma/J=2$.

\begin{figure}[htbp]
	\begin{center}
		\includegraphics[width=\columnwidth, angle=0]{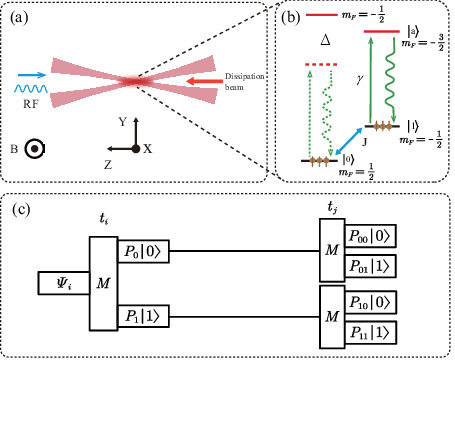}
		\caption{
			The experimental setup for the non-Hermitian LGI. (a) The schematic illustrates the direction of the magnetic field, RF pulse, and dissipative light, along with the corresponding axes in the Bloch sphere. (b) The blue arrow represents the coupling between the two levels generated by the RF pulse with strength J, while the green solid arrow indicates the dissipation of spin-dependent atomic loss due to a resonant optical beam with strength $\gamma$. These two quantities enable the realization of the non-Hermitian Hamiltonian in Eq.(1). The resonant light from state $\ket{1}$ to state $\ket{a}$ will also excite atoms in state $\ket{0}$ to their corresponding excited state with a detuning of $\Delta=75.655MHz$, as shown by the dashed arrows. The dotted and solid wavy arrows represent the spontaneous emission from their excited sates to the ground states. These three quantities can be used to describe the mechanisms of decoherence in the non-Hermitian dynamical evolution process.(details are discussed in Discussion) (c) The schemes of getting three-times correlation parameter $K_3$ by testing two-time correlation function for three times. The experiment includes measuring observable values at times of $t_i$, $t_j$($i=1,2;j=2,3;i\neq j$). Here, $\ket{\psi_i}$ is the target state, and $t_j-t_i=(j-i)\tau$ by $t_1=0$.
		}
	\end{center}
\end{figure}

LGI serves as a method for testing the coherent effects of individual systems at the macroscopic scale by examining correlations among separated moments. The typical three-time correlation function $K_3$ is defined as ~\cite{emary2013leggett}:
\begin{equation}
	K_3=C_{32}+C_{21}-C_{31}
\end{equation}
where $C_{ij}(i,j=1,2,3; i>j)$ is the two-time correlation function between $t_j$ and $t_i$ following the form:
\begin{equation}
 C_{ij}=\sum_{Q_{i}, Q_{j}= \pm 1} Q_{i} Q_{j} P_{i j}\left(Q_{i}, Q_{j}\right)
\end{equation}
The specific calculation expressions are described in the Appendix B.  Here we take the observable $\sigma_y$ with the eigenvectors $\ket{+}=\sqrt{1/2}(i \ket{0}+\ket{1})$ and $\ket{-}=\sqrt{1/2}(-i \ket{0}+\ket{1})$. $Q_i$ and $Q_j$ are the measurement values of the states at time $t_i$ and $t_j$, which should be the eigenvalues $+1$ and $-1$ of the $\sigma_y$. $P_{ij}(Q_i,Q_j)$ is the probability of getting $Q_i$ at $t_i$ and $Q_j$ at $t_j$.

For classical systems, $-3\le K_3\le 1$, and in a two-level quantum system, the upper bound of $K_3$ can exceed 1 and reach 1.5~\cite{budroni2014temporal,emary2013decoherence} with invariant lower bound. Further in non-hermitian quantum system, the maximum violation of LGI can approach 3.

As shown in Fig. 1(a), we define the direction of the main magnetic field as the $\boldsymbol{Z}$ axis, with the direction of the RF field perpendicular to the $\boldsymbol{Z}$ axis, defined as the $\boldsymbol{X}$axis. Fig. 1(c) illustrates the experimental scheme. We first dissipate the atoms in state $\ket{1}$ to obtain a pure state $\ket{0}$ (details are discussed in Appendix A). Then, we apply a $\pi/2$ pulse to create a coherent state $\ket{+}$ as our initial state. Next, we allow the prepared state to undergo non-unitary evolution under the hamiltonian described in Eq.(1) for a certain duration. Finally, another $\pi/2$ pulse is applied to rotate the projection of the Bloch state on the $\boldsymbol{Y}$ axis around the $\boldsymbol{X}$ axis to the $\boldsymbol{Z}$ axis, enabling us to measure the observable $\sigma_y$ by projecting onto the $\boldsymbol{Z}$ axis.
 
 
 However, with a large dissipation strength $\gamma$, the atom number decays rapidly, significantly reducing the signal-to-noise ratio of the data. To address this, we divide the total evolution time $T$ into $n$ segments. We then prepare the initial state for each segment and allow them to evolve for $\tau =T/n$ individually~\cite{lu2023}. Finally, we recombine the results of these individual evolutions into a complete cycle. We refer to $\tau$ as the effective evolution time. This approach enables effective mapping of the entire process, while the experimental results are detailed in Appendix B. It is noted that, $\tau$  is a crucial factor influencing the evolution, as shown in the results section.

\section{LGI violation enhance due to non-Hermitian interactions}

We observe violations of the LGI during the Hermitian evolution of ultracold Fermi gases. Moreover, we detect an enhancement in the violation of the LGI in non-Hermitian evolution after introducing dissipation.

\begin{figure}[htbp]
	\begin{center}
		\includegraphics[width=\columnwidth, angle=0]{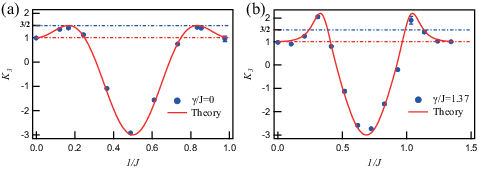}
		\caption{
			The results of $K_3$. (a) The result under Hermitian condition ($\gamma/J$ = 0). (b) The measurement result of $K_3$ under non-Hermitian condition, the $\gamma/J$ are 1.37. The red solid lines are the theoretical results, and the blue markers are the measured results. The red dash line is the upper bound 1 in the classical system, the blue dash line is the upper bound 1.5 in the Hermitian quantum system. The $\boldsymbol{X}$ axis in figure represents the time with the unit of second, and we normalized it to the Rabi cycle $1/J$ for convenience.
		}
	\end{center}
\end{figure}


Fig. 2 illustrates the time evolution of the $K_3$ under both Hermitian and non-Hermitian conditions. We begin with an initial state of $\ket{+}$. As the duration of the RF field varies, the time-correlation paramete $K_3$, shown in Fig. 2(a), reaches a maximum value of 1.43 at approximately 0.2 Rabi cycles (0.22 ms), which deviates from the classical limit by 10 standard deviations. This result indicates a violation of the LG inequality due to coherence during the coupling process, highlighting the quantum effects of the evolution at a macroscopic scale.

When adding dissipation beam synchronized with the RF coupling field, we get a dissipation non-Hermitian system. We adopt a segmented experiment approach mentioned before. We take the effective evolution time as 50 $\mu s$. By integrate the result of each segment we obtain the time evolution curve in the non-Hermitian system, as presented in Fig.2(b). At about 0.4 Rabi cycles(0.37 $ms$), the maximum value of $K_3$ reaches 2.06, deviating from the hermitian limit by 11 standard deviations. The results demonstrate that in a non-Hermitian system, the upper bound of the LGI parameter $K_3$ surpasses the limit of the Hermitian case 1.5. This phenomenon arises from the non-uniform speed of quantum state evolution in the non-Hermitian process~\cite{lu2024realizing}.

\section{LGI violation weakens due to decoherence in $\gamma/J$ domain}

During the non-Hermitian evolution, we observed a weakening in the violation of the LGI caused due to decoherence as $\gamma/J$ increases.

\begin{figure}[htbp]
	\begin{center}
		\includegraphics[width=\columnwidth, angle=0]{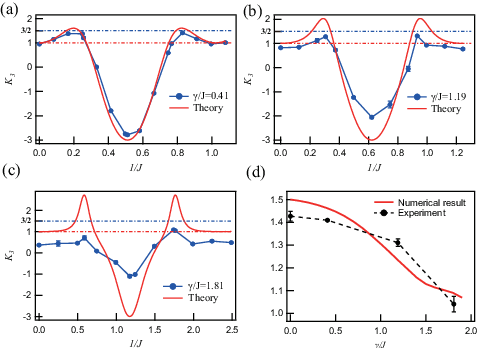}
		\caption{
			Experimental results for the $K_3$ with different $\gamma/J$ (a)-(c) Measurement results of $K_3$ under non-Hermitian condition, with the $\gamma/J$ being 0.42 (b), 1.19 (c), and 1.81 (d), respectively. The red solid line in (a)-(c) are the theoretical prediction with non-Hermitian Hamiltonian shown in Eq.(1). (d) The black dots are the maximum value of $K_3$ in different $\gamma/J$ from the results in Fig.2(a) and Fig.3(a)-(c) with $J=2\pi\times 414.91Hz$. The red line represents the numerical results derived from integrating a basic decoherence model with experimental procedures(as elaborated in DISCUSSION section).
		}
	\end{center}
\end{figure}

In our experiments, we found that when the  effect evolution time of the quantum state is 50$\mu s$, the violation of LGI intensifies as the dissipation strength increases, as shown in Fig.2(b). This is manifested by the $K_3$ exceeding the Hermitian limit of 1.5. However, when we choose the effect evolution time to be 500$\mu s$, we observe a gradual weakening of the violation of LGI with increasing dissipation strength. As depicted in Figure 3, for instance, in Fig.3(a) with $\gamma/J=0.41$, the experimentally measured maximum value of $K_3$ is $1.41$, differing from the theoretical maximum value of 1.51 for the non-Hermitian case by 10 standard deviations. In Fig.3(b) with $\gamma/J=1.19$, $K_3$ reaches a maximum value of $1.31$, differing from the theoretical value of 2.08 by 20 standard deviations. In Fig.3(c) with $\gamma/J=1.81$, $K_3$ attains a maximum value of $1.04$, deviating from the theoretical value of 2.8 by 45 standard deviations, and only differing from the classical boundary of 1 by one standard deviation. 

These results indicate that as the dissipation strength increases, theoretically predicted upper bounds for $K_3$ should exhibit an increasing violation, but experimental results instead show a decreasing trend. The experimental results under non-Hermitian conditions diverge significantly from theoretical predictions, even approach classical domain in the longer evolution time. Moreover, when $\gamma/J$ approaches the EP point, experimental results no longer indicate a violation of the classical limit of 1, nor can they signify the entire evolution process as quantum. We even have reason to believe that with increasing dissipation strength, the system's evolution undergoes a transition from quantum to classical.

\section{LGI violation weakens due to decoherence in time domain}

We observe that increasing dissipation strength weakens the LGI violation at a fixed evolution time. Similarly, at a fixed dissipation strength, extending the evolution time also diminishes the LGI violation.

\begin{figure}[htbp]
	\begin{center}
		\includegraphics[width=\columnwidth, angle=0]{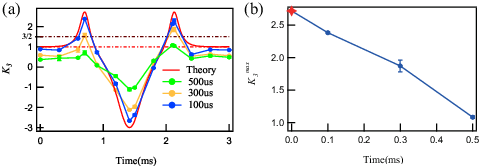}
		\caption{
			Experimental results for the $K_3$ with different effect evolution time. (a) Red line is the theoretical prediction with non-Hermitian Hamiltonian shown in Eq.(1). Blue, yellow and green dots are the experiment results when the effect evolution time is 100 $\mu s$, 300 $\mu s$ and 500 $\mu s$, respectively. (b) The red rhombus is the maximum value of $K_3$ predict by the theory. The blue dots are the maximum value of $K_3$ in different effect evolution time from the results in (a) with the same $J=2\pi\times 414.91Hz$ and the same $\gamma=1.81\times J$
		}
	\end{center}
\end{figure}

To further investigate this transition, we selected dissipation strengths close to the EP point($\gamma/J=1.81$). As we gradually increased the effect evolution time, the upper bound of $K_3$ gradually decreased. Fig.4(a) shows the evolution curves of $K_3$ at different evolution times, while Fig.4(b) displays the curve of the maximum value of $K_3$ as a function of the effect evolution time $\tau$. We selected effect evolution times of 100 $\mu s$, 300 $\mu s$, and 500 $\mu s$. When the effect evolution time reached 500 $\mu s$, the upper bound of $K_3$ degraded from the theoretically predicted 2.61 to $1.04$, exceeding the classical limit of 1 by only one standard deviation, indicating the disappearance of quantum behavior. In other words, as the evolution time increased, the system undergoes a transition from quantum to classical.

This transition is believed to be caused by decoherence. In the experiment, we excite one of the two ground states to an excited state using a resonant light beam. Dissipation typically accompanies decoherence effects~\cite{schlosshauer2019quantum}, which can be described using the open system master equation. As the timescale of evolution gradually increase, the quantum jump terms become significant. The longer the evolution time, the stronger the system's decoherence. When this evolution time exceeds a critical value, the system transitions from quantum to classical evolution. This aligns with the conclusions drawn from our experiment.

\section{disscussion}

When decoherence effects are absent, the primary reason for the enhancement of LGI violation during non-Hermitian evolution is the uneven speed of evolution from quantum state $\ket{1}$ to $\ket{0}$ and back to $\ket{1}$. This is clearly discussed in the theoretical discourse in ~\cite{varma2023extreme} and has been well experimentally verified in~\cite{lu2023}.

From the perspective of the master equation, 
\begin{equation}
\frac{\partial \rho}{\partial t}=-i\left[H_{c}, \rho\right]+\sum_{k=a,d}\left[L_{k} \rho L_{k}^{\dagger}-\frac{1}{2}\left\{L_{k}^{\dagger} L_{k}, \rho\right\}\right]
\end{equation}
where $L_{a}=\sqrt{2 \gamma}|a\rangle\langle 1|$.
For a three-level system, the RF field acts on states $\ket{0}$ and $\ket{1}$, coupling the two levels. Resonant dissipative light acts on state $\ket{1}$, dissipating atoms from state $\ket{1}$ to the excited state, denoted as state $\ket{a}$. The quantum jump term can be written as $|a\rangle\langle 1|\rho|1\rangle\langle a|=\rho_{11}|a\rangle\langle a|$(acting only in excited state $\ket{a}$), which is irrelevant to the subsystem of levels $\ket{0}$ and $\ket{1}$ and can be ignored. By solving the master equation for the density matrix, we can obtain an analytical solution for non-Hermitian evolution, and consequently, the uneven evolution pattern of the Bloch $z$ component, as shown in Figure 5.

In the experiment, the relative orientation of our main magnetic field and the RF magnetic field determines that our coupling manifests as counterclockwise evolution on the Bloch sphere. In other words, in the $x-z$ plane, the quantum state evolves from $z+$ through $x-$ to $z-$, then through $x+$ back to $z+$. Therefore, when we activate the coupling RF field and dissipate the $z+$ state, during the transition from $z+$ to $z-$, coupling and dissipation compete, resulting in a slowdown of the evolution rate. However, during the evolution from $z-$ back to $z+$, the coupling and dissipation have the same direction of speed, resulting in an overall acceleration of evolution. This uneven evolution rate leads to nonlinear interactions between the dissipative two-level system, as described by Ref.~\cite{varma2023extreme,lu2023,zhan2023}

\begin{figure}[htbp]
	\begin{center}
		\includegraphics[width=\columnwidth, angle=0]{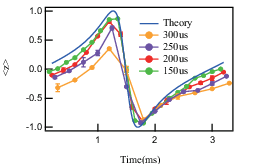}
		\caption{
			Blue line is the $z$ component on Bloch sphere predict by the non-Hermitian Hamiltonian shown in Eq(1). Green, red, purple and yellow dots are the experiment results when the effect evolution time is 100 $\mu s$, 200 $\mu s$ ,250 $\mu s$, and 300 $\mu s$, respectively. 
		}
	\end{center}
\end{figure}

However, dissipative light is always accompanied by the presence of decoherence effects. We describe decoherence effects using the master equation by adding $L_{d}=\gamma_{d}|1\rangle\langle 0|$~\cite{chen2022decoherence} representing a spontaneous transition from $\ket{0}$ to $\ket{1}$. Actually, in $^{6}$Li atomic energy levels, there is no spontaneous emission between the two ground-state levels that constitute the pseudospin. However, when dissipative light that excites transitions from the ground state $\ket{F=1/2, m_F=1/2}$($\ket{1}$) to the excited state $\ket{F=5/2, m_F=-1/2}$ is introduced, the ground-state levels become coupled to the excited states. This process can be interpreted as an interaction between the ground state $\ket{1}$ and the environmental excited state $\ket{F=5/2, m_F=-1/2}$. Additionally, the ground state $\ket{F=1/2, m_F=-1/2}$($\ket{0}$) is coupled to the excited state $\ket{F=5/2, m_F=-3/2}$ with a detuning of 75.655 MHz(527.3G). Since a larger detuning results in a weaker coupling strength, the interaction between the ground state $\ket{0}$ and the excited state $\ket{F=5/2, m_F=-3/2}$ is weaker than that between the ground state $\ket{1}$ and the excited state $\ket{F=5/2, m_F=-1/2}$. Consequently, this process can be effectively viewed as a spontaneous emission process from the ground state $\ket{0}$ to the ground state $\ket{1}$. Furthermore, as the intensity of the dissipative light increases, the interaction strength also increases, leading to stronger decoherence effects.


This kind of transition is always seen as one of the effect of decoherence, $\gamma_{d}$ serving as the decoherence strength. Similarly, for a three-level system, as atoms in state $\ket{1}$ are excited to state a by a resonant dissipative light beam, there is also a spontaneous emission process from state $\ket{1}$ to state $\ket{0}$. Therefore, the $L$ operator in the master equation is rewritten as $L=L_{a}+L_{d}$. As mentioned before $\gamma$ from $L_{a}$ acting solely on the excited state $\ket{a}$ does not affect the evolution of state $\ket{1}$ and $\ket{0}$, nor does it induce decoherence in this two-level subsystem. However, when we incorporate $L_{d}$ into Eq.(4) the quantum jump term caused by $L_{d}$ can be expressed as $|1\rangle\langle 0|\rho|0\rangle\langle 1|=\rho_{00}|1\rangle\langle 1|$. This indicates that $\gamma_{d}$ from $L_{d}$ acts directly on state $\ket{1}$, leading to decoherence between $\ket{0}$ and $\ket{1}$. Considering that higher light intensity results in stronger decoherence, we assume a simple model where $\gamma_{d}=\alpha\gamma_{a}$, indicating that as dissipation strength increases, decoherence strength also increases linearly. By substituting this relationship into the main Eq.(4) and fitting the data points from Fig.3(d), we obtain a coefficient $\alpha=0.58$. The red line in Fig.3(d) represents the numerical fit curve, which agrees with the experimental result fairly well.


Additionally, we conducted numerical simulations for different evolution times under the same dissipation strength. It is observed that as the evolution time increases, the efficiency of the transition from state $\ket{0}$ to state $\ket{1}$ gradually decreases(as shown in Fig.5), which is a manifestation of decoherence. In other words, as the evolution time lengthens, the degree of decoherence increases. With the increase in decoherence, we found through time-domain correlation function tests that the violation of the LGI gradually weakens until it no longer violates and transitions to classical evolution.

In summary, when decoherence effects are absent, non-Hermitian evolution leads to an enhancement of LGI violation, while the presence of decoherence effects weakens LGI violation. There exists a competitive relationship between the two mechanisms. When decoherence effects are strong enough, the system's evolution will undergo a transition from quantum to classical. Such a competing mechanism let us better understand the QCT properties in the non-Hermitian system. LGI also provides a useful method to probe the decoherence effects using the time domain quantum correlation.

\section*{Acknowledgements}

We thank Drs.Xinxin Rao and Pengfei Lu for the helpful discussions. This work is supported by the National Key Research and Development Program under Grant No.2022YFC2204402, the National Natural Science Foundation of China under Grant No.12174458, Guangdong Science and Technology project under Grant No.20220505020011, the Fundamental Research Funds for the Central Universities (Sun Yat-Sen University, 2021qntd28 and 2023lgbj020), SYSU Key Project of Advanced Research, Shenzhen Science and Technology Program under Grant No.JCYJ20220818102003006, and Shenzhen Science and Technology Program under Grant No.2021Szvup172



\appendix
\section{Experimental setup}

After collecting and precooling the atoms using a magneto-optical trap (MOT), they are loaded into a far-detuned optical dipole trap. This trap is formed by two Gaussian beams with 37 $\mu m$ waists, a 12-degree angle between them, and a 1064 $nm$ wavelength, sourced from an IPG Photonics YLR-100-1064-LP continuous fiber laser. Subsequently, the magnetic field is swept to 330Gs, and a RF pulse is applied to achieve a 50:50 mixture of $^{6}$Li fermions in the lowest hyperfine states. A 3.2 s ramp-down of the optical trap is then executed for evaporative cooling, reaching a target trap depth of 0.26 $\mu K$ (with a full trap depth of 5.6 $mK$). After the cooling process, the magnetic field is precisely adjusted to 527.3 Gs, rendering the s-wave scattering lengths for $\ket{0}$ and $\ket{1}$ to be zero, thus obtaining a non-interacting Fermi gas. In order to obtain a pure ensemble in the coherent $\ket{0}$, we dissipate all atoms in the $\ket{1}$. Typically, we have about 2$\times10^{5}$ atoms in pure $\ket{0}$ states at a temperature of 0.26 $\mu K$ and a $T/T_f$ of about 0.3, where $T_f$ is the Fermi temperature.

\section{Calculations and experimental results of two-times correlation functions}

\begin{figure}[htbp]
	\begin{center}
		\includegraphics[width=\columnwidth, angle=0]{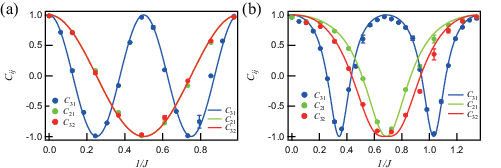}
		\caption{
			Experimental results of correlation functions $C_{ij}$.(a) The measurement result of $C_{ij}$ under Hermitian condition ($\gamma/J$ = 0). (b) The measurement result of $C_{ij}$ under nonHermitian condition, the ratios of $\gamma/J$ is 1.37. The solid line in (a)(b) are the theoretical results, and the markers are the experimental measured results with blue($C_{31}$), green($C_{21}$) and red($C_{32}$)
		}
	\end{center}
\end{figure}

The fundamental premise underpinning the application of the LGI to discern classical from quantum behavior lies in its reliance on two classical postulates. In classical systems, measurements yield deterministic outcomes at any given moment, and these measurements exert no influence on the subsequent evolution of the system. Conversely, in quantum systems, the act of measurement inherently disturbs the system's evolution. Thus, we conduct three joint measurements at distinct time points, each assessing the system's state at two of these time points. A deviation from the expected evolution due to measurement signifies a violation of LGI, characterized by the upper bound of $K_3$ surpassing the classical limit of 1. Through eq (2) and (3), we obtain analytical expressions for the the two-times correlation function $c_{ij}$:

\begin{equation}
\begin{aligned}
	C_{21}&=\frac{\gamma+J \cos (2 \tau \chi)}{J+\gamma \cos (2 \tau \chi)},\\
	C_{31}&=\frac{\gamma+J \cos (2 \tau \chi)}{J+\gamma \cos (2 \tau \chi)},\\
	C_{32}&=\frac{J \gamma^{2}+J\left(J^{2}+J \gamma-\gamma^{2}\right) \cos (2 \tau \chi)}{[J-\gamma \cos (2 \tau \chi)][J+\gamma \cos (2 \tau \chi)]^{2}}\\
	&-\frac{\gamma \cos ^{2}(2 \tau \chi)\left[-J^{2}+J \gamma+\gamma^{2}+J^{2} \cos (2 \tau \chi)\right]}{[J-\gamma \cos (2 \tau \chi)][J+\gamma \cos (2 \tau \chi)]^{2}}.
\end{aligned}
\end{equation}

Here, the time intervals are defined as $t_3-t_2=t_2-t_1=\tau$ and $\chi=\sqrt{J^2-\gamma^2/4}$.The experimentally measured two-times correlation function $C_{ij}$ with short effect evolution time(ignore the decoherence) are depicted in Fig.6, which agree well with the theoretical predictions.\cite{zibrov1996experimental}


%

\end{document}